\documentstyle[12pt]{article}

\newcommand{\be}{\begin{eqnarray}}
\newcommand{\en}{\end{eqnarray}}

\begin{document}
\begin{titlepage}
\begin{flushright}
KOBE-TH-97-03  \\
hep-th/yymmnn
\end{flushright}

\begin{center}  
\vskip 0.3truein

{\Large\bf {Duality and Superconvergence Relation }} \\
\vskip0.07truein
{\Large\bf {in Supersymmetric Gauge Theories}}

\vskip1.0truein

{Motoi Tachibana} 
\footnote{E-mail: tatibana@oct.phys.kobe-u.ac.jp}

\vskip0.2truein

{\it Graduate School of Science and Technology}

{\it Kobe University, Rokkodai, Nada, Kobe 657, Japan}

\end{center}
\vskip0.5truein \centerline{\bf Abstract} \vskip0.13truein

We investigate the phase structures of various $N=1$
supersymmetric gauge theories including even the exceptional
gauge group from the viewpoint
of superconvergence of the gauge field propagator.
Especially we analyze in detail whether a new type of
duality recently discovered by Oehme in $SU(N_c)$ gauge theory
coupled to fundamental matter fields
can be found in more general gauge theories with more general
matter representations or not. The result is that in the cases
of theories including matter fields 
in {\it only} the fundamental representation, Oehme's
duality holds but otherwise it does not. In the former case,
superconvergence relation might give good criterion to
describe the interacting non-Abelian Coulomb phase without
using some information from dual magnetic theory.

\end{titlepage}
\newpage
\baselineskip 20 pt
\pagestyle{plain}
\vskip0.2truein

\setcounter{equation}{0}

\vskip0.2truein

\section{Introduction}

Quark confinement is one of the most mysterious properties
in quantum field theory. In spite of its obvious existence 
in experiment, the mechanism of confinement has not been
elucidated yet. Of course there have ever been many challenges
to understand this phenomena. Lattice formulation of QCD,
originally proposed by Wilson \cite{wil}, is one of them.
On the other hand, chiral symmetry breaking is another problem
which must be solved. In non-supersymmetric gauge theory,
it is considered that these two phenomena, confinement and
chiral symmetry breaking, are deeply connected and occur
simultaneously in a certain QCD parameter region (strong
gauge coupling or small number of quark flavors etc.).

 Our main interest here is how the phase structure of QCD
changes as the number of quark flavors increases. Naively
we expect the following picture: when the number is small,
theory is in confinement phase due to its asymptotic freedom
property. As the number of flavors is increasing,
quarks are deconfined  and chiral symmetry is restored 
before theory becomes asymptotically non-free.

In the intermediate region corresponding to
no confinement but asymptotic freedom, 
we can realize the scale invariant theory
because it is expected 
there exists a non-trivial IR fixed point in this region.
The critical value of $N_f$ in $SU(3)$ QCD
where quarks are deconfined
and chiral symmetry is restored
has been evaluated by many authors.
Banks and Zaks first pointed out the existence of such a 
fixed point \cite{bz}. They evaluated the first two coefficients
of perturbatively expanded $\beta$-function which are
both gauge and renormalization scheme independent \cite{gro} .
They showed $N^{crit}_f
= 8.05$. Also $N^{crit}_f = 7$ has been obtained 
in lattice QCD calculation \cite{tsu}. 
On the other hand, Oehme and Zimmermann expected
$N^{crit}_f = 10$ using superconvergence relation
they called \cite{oz}. 
In this relation, the anomalous dimension of 
the gauge field as well as the $\beta$-function plays an
important role. In this way, all the values of $N^{crit}_f$
in different approaches do not coincide each other.

 Recently Oehme has applied his superconvergence argument
to $N=1$ supersymmetric $SU(N_c)$ gauge theory and 
compared with already known results \cite{oeh} . 
The phase structure
of this theory has been already investigated in detail
with the help of so-called \lq \lq electric-magnetic''
duality and holomorphy by Seiberg et al. \cite{sei} .
Especially Seiberg has insisted on the existence of
an interval corresponding to interacting non-Abelian
Coulomb phase or conformal window where the theory
becomes scale invariant. Consequently Oehme showed
quantitative agreement between his argument and the results
from Seiberg's duality. Moreover he also found the
important relationship between original electric SUSY
theory and dual magnetic one, which might be interpreted
as a new type of duality. Since superconvergence arguments
can apply for {\it both SUSY and non-SUSY} theories,
the comparison with exact results by Seiberg et al.
is very significant.

In this paper, we apply his method to various supersymmetric
gauge theories with other gauge groups and other matter
contents and check whether the relation he found holds or not
in those cases.
In section 2, we review the concept of superconvergence
of the gauge field propagator in non-supersymmetric case.
In section 3, we extend the method in the previous section to
the supersymmetric cases and check the Oehme's duality.
Section 4 is devoted to summary and discussions. In Appendices,
some basic equations are followed.

\section{Superconvergence Relation}

 Here we shall consider the asymptotic behavior of the gluon
propagator at large momentum with the help of the renormalization
group (RG) analysis.

 First of all, following Oehme and Zimmermann \cite{oz},
we introduce the operator such as

\begin{equation}
A^{a}_{\mu\nu} \equiv \partial_{\mu}A^{a}_{\nu}-\partial_{\nu}A^{a}_{\mu},
\label{trans}
\end{equation}
where $A^{a}_{\mu}(\mu=1,\cdots,4,$ and $a=1,\cdots,N^2_c -1)$ is the 
$SU(N_c)$ gauge field whose two point function is generally given by

\begin{eqnarray}
& & <0|TA^{a}_{\mu}(x)A^{b}_{\nu}(y)|0>  \nonumber \\
&=& \int d^4k e^{-ip \cdot (x-y)} \frac{\delta^{ab}}{i}
  [(\delta_{\mu \nu} - \frac{k_{\mu}k_{\nu}}{k^2})D(k^2)
  + \alpha \frac{k_{\mu}k_{\nu}}{k^4}].
\label{kugo}
\end {eqnarray}
$\alpha$ is the gauge parameter.

Using (\ref{trans}) and (\ref{kugo}), we will obtain

\begin{eqnarray}
& &<0|TA^{a}_{\mu\nu}(x)A^{b}_{\rho\sigma}(y)|0> = \nonumber \\
& &\int d^4k e^{-ip \cdot (x-y)} \frac{\delta^{ab}}{i}
  (k_{\mu}k_{\rho}\delta_{\nu\sigma} - k_{\mu}k_{\sigma}\delta_{\nu\rho}
- k_{\nu}k_{\rho}\delta_{\mu\sigma} + k_{\nu}k_{\sigma}\delta_{\mu\rho})
D(k^2).
\label{kugo2}
\end {eqnarray}
Scalar function $D(k^2)$ is called $\lq \lq$ transverse gluon
structure function". Below we shall restrict our consideration
to the asymptotic behavior of this function $D(k^2)$ at large momentum.

For that purpose, first we give the normalization condition 
for $D(k^2)$ as follows:

\begin{equation}
k^2 D(k^2) = 1  \qquad \qquad at \quad k^2 = {\mu}^2,
\label{normal}
\end{equation}
where $\mu$ is the normalization point. Then we can write
the structure function in the form

\begin{equation}
D = D(k^2, g, {\mu}^2).
\label{dfunc}
\end{equation}
$g$ is the gauge coupling constant. 

Now, in convenience, let us introduce the dimensionless function
$R$ defined by

\begin{equation}
R \equiv k^2 D(k^2, g, {\mu}^2) = R(\frac{k^2}{{\mu}^2}, g).
\label{rfunc}
\end{equation}

Then Callan-Symanzik equation (RG equation) for the $R$-function
in the Landau gauge ($\alpha=0$) is given as

\begin{equation}
u\frac{\partial R(u,g)}{\partial u}
= \beta (g^2)\frac{\partial R(u,g)}{\partial g^2}
 +\gamma (g^2)R(u,g),
\label{cgeq}
\end{equation}
where $u \equiv \frac{k^2}{{\mu}^2}$ and $\beta (g^2)$ and
$\gamma (g^2)$ are the $\beta$-function and the anomalous
dimension of the gluon field, respectively. In the region
of small gauge coupling constant (i.e., at large momentum),
they are of the form:

\begin{eqnarray}
\beta(g^2) &=& g^4({\beta}_0 +{\beta}_1 g^2 + \cdots), \nonumber \\
\gamma (g^2) &=& g^2({\gamma}_0 +{\gamma}_1 g^2 \cdots),
\label{perturb}
\end{eqnarray}
where
\begin{eqnarray}
{\beta}_0 &=& -\frac{1}{16{\pi}^2}(\frac{11}{3}N_c-\frac{2}{3}N_f), 
\nonumber \\
{\gamma}_0 &=& -\frac{1}{16{\pi}^2}(\frac{11}{6}N_c-\frac{4}{3}N_f).
\label{coeff}
\end{eqnarray}
$N_f$ denotes the number of quark flavors.

 Oehme and Zimmermann solved the eq.(\ref{cgeq}) in the following
form (See Appendix A):

\begin{equation}
R(\frac{k^2}{{\mu}^2},g) = 
R(1,Q) exp[\int^{Q^2}_{g^2}dx \gamma(x) {\beta}^{-1}(x)].
\label{sol}
\end{equation}
Here the effective RG-invariant coupling constant 
$Q(\frac{k^2}{{\mu}^2},g)$ is defined through the equation

\begin{equation}
u \frac{\partial Q^2(u,g)}{\partial u}
= \beta (g^2) \frac{\partial Q^2(u,g)}{\partial g^2}.
\label{qeq}
\end{equation}
For large momentum, we can show that

\begin{equation}
Q^2(\frac{k^2}{{\mu}^2},g)
\approx -\frac{1}{{\beta}_0 \ln{\frac{k^2}{{\mu}^2}}}.
\label{qsol}
\end{equation}
${\beta}_0$ is the first coefficient of the $\beta$-function
defined in eq.(9).

 Now we would like to obtain the asymptotic behavior of the
$R$-function given by eq.(\ref{sol}). By substituting (\ref{perturb})
and (\ref{coeff}) into (\ref{sol}), we can get the following result
(see Appendix B) 
\begin{eqnarray}
R(\frac{k^2}{{\mu}^2},g) &=& 
(\frac{Q^2}{g^2})^{\frac{{\gamma}_0}{{\beta}_0}}
exp[\int^{Q^2}_{g^2}dx \tau(x)], \nonumber \\
&\approx& C_V(\ln{\frac{k^2}{{\mu}^2}})^{-\frac{{\gamma}_0}{{\beta}_0}}
\label{asymp}
\end{eqnarray}
where
\begin{equation}
C_V = (g^2 |{\beta}_0|)^{-\frac{{\gamma}_0}{{\beta}_0}}
exp[\int^{Q^2}_{g^2}dx \tau(x)].
\label{regular}
\end{equation}
Here the function $\tau (x)$ is the regular part of 
$\frac{\gamma(x)}{\beta(x)}$ at $x=0$.

 Thus leading asymptotic behavior of the $R$-function at
large momentum was determined. Similarly, for the $D$-function,
we find

\begin{equation}
D_{asymp}(k^2) \approx
C_V k^{-2} (\ln{\frac{k^2}{{\mu}^2}})^{-\frac{{\gamma}_0}{{\beta}_0}}.
\label{asymp2}
\end{equation}

We can conclude here
that the asymptotic behavior of the transverse gluon propagator
drastically changes due to the sign of ${\gamma}_0$ (or
equivalently the sign of $\frac{{\gamma}_0}{{\beta}_0}$).
If $\gamma_0 < 0$, it converges. While if $\gamma_0 > 0$, it diverges
where we have assumed the asymptotic freedom of the theory
which means $\beta_0 <0$.

Let us consider such a  case that ${\beta}_0$ and
${\gamma}_0$ are both negative, i.e., the ratio
$\frac{{\gamma}_0}{{\beta}_0}$ is positive. For the scalar
part of the transverse gluon propagator $D(k^2)$, we can apply
the Lehmann representation as

\begin{equation}
D(k^2) = 
\int^{\infty}_{0}dm^2 \frac{\rho(m^2)}{m^2 - k^2},
\label{lehmann}
\end{equation}
where $\rho(k^2)$ is called the spectral function and is 
given as the absorptive part of the $D$-function:

\begin{equation}
\pi\rho(k^2) = ImD(k^2) = k^{-2}ImR(k^2).
\label{spectral}
\end{equation}
Therefore, for the limit $k^2 \to -\infty$, we find

\begin{equation}
\rho_{asymp} \approx -\frac{{\gamma}_0}{{\beta}_0} C_V k^{-2}
(\ln{\frac{k^2}{{\mu}^2}})^{-\frac{{\gamma}_0}{{\beta}_0}-1}.
\label{spectral2}
\end{equation}
Combining with (\ref{asymp2}),(\ref{lehmann}) and 
(\ref{spectral2}), a kind of sum rule, called superconvergence
relation can be obtained (see Appendix C)

\begin{equation}
\int^{\infty}_{0}dm^2 \rho(m^2) = 0
\qquad \qquad for \quad \frac{\gamma_0}{\beta_0} > 0.
\label{screlation}
\end{equation}

 We can show that superconvergence
relation obtained here gives some circumstantial evidence for
color confinement. In \cite{nisi}, it has been said that 
superconvergence relation connects with various interpretations
of color confinement, such as metric cancellation, the bag
model picture and the area law behavior of the Wilson loop
in lattice QCD. 

On the contrary, in the region where
superconvergence relation does not hold
(i.e., $\beta_0 < 0$ and $\gamma_0 > 0$), 
quarks are deconfined and chiral symmetry restored.
Moreover it is expected that there exist a non-trivial IR fixed
point in this region and the gauge coupling cannot
be too strong to confine quarks and to occur quark condensate.
In the following section, we will restrict our considerations
to this interval in supersymmetric gauge theories.
\section{Duality and Superconvergence Relation in
         various N = 1 SUSY Gauge Theories}

In the previous section, we discussed superconvergence
relation (a kind of sum rule for the spectral function of
the transverse gluon propagator) and 
commented its relation with color
confinement. There the essential point was that there existed
the region in which we have asymptotic freedom of the theory
but no confinement. In such a region, theory may have a
non-trivial infrared fixed point at non-vanishing value of
the gauge coupling.

 On the other hand, Oehme has recently applied the method
of superconvergence relation to $N$=1 supersymmetric
gauge theory ($SU(N_c)$ gauge theory with fundamental chiral
supermultiplet of $N_f$ flavors) \cite{oeh}
and compared the result
with the one Seiberg et al.  
have already obtained using holomorphy and 
\lq \lq electric-magnetic'' duality \cite{sei}.
Oehme has insisted in \cite{oeh} that his result 
is in quantitative agreement
with Seiberg's duality argument. Moreover he showed that
there was an interesting relationship between the coefficients
$\beta_0$ and $\gamma_0$ (defined in the previous section)
of the original (electric) theory and those of the dual
(magnetic) theory. This may be considered as a new type 
of duality.

 Thus in this section, we first review the analysis by Oehme in
\cite{oeh} and then try to apply his method to other 
models of $N=1$ SUSY gauge theories, 
which have different gauge groups and different
matter representations, to check Oehme's duality. That is the 
main purpose of this paper.

\vskip0.2truein
{\bf (I) Oehme's analysis in $N$=1 $SU(N_c)$ gauge theory
with massless matter fields in 
the fundamental representation of $N_f$ flavors}

In this case, the one loop coefficients $\beta_0$ and $\gamma_0$
are given by
\begin{eqnarray}
{\beta}_0 &=& -\frac{1}{16{\pi}^2}(3N_c - N_f), 
\nonumber \\
{\gamma}_0 &=& -\frac{1}{16{\pi}^2}(\frac{3}{2}N_c - N_f).
\label{coeffi2}
\end{eqnarray} 
If we demand $\beta_0 < 0$ (asymptotic freedom) and 
$\gamma_0 > 0$ (deconfinement), then we will have the 
following result:
\begin{equation}
\frac{3}{2}N_c < N_f < 3N_c.
\label{conformal}
\end{equation}

This interval is just corresponding to the one Seiberg called
interacting non-Abelian Coulomb phase or conformal window.
The theory in this interval has a non-trivial infrared fixed
point and becomes scale invariant. 

 On the other hand, as has been well known, we have the dual
magnetic description for the original electric theory in the
region (\ref{conformal}). 
The gauge group of dual theory is
$G_{dual} = SU(N_f-N_c)$ with $N^{dual}_f=N_f$ flavors of
magnetic chiral superfields and a certain number of singlet
massless superfields. In this dual theory, one-loop 
coefficients $\beta^{d}_0$ and $\gamma^{d}_0$ are
\begin{eqnarray}
{\beta}^{d}_0 &=& -\frac{1}{16{\pi}^2}(2N_f -3N_c ), 
\nonumber \\
{\gamma}^{d}_0 &=& 
           -\frac{1}{16{\pi}^2}(\frac{1}{2}N_f -\frac{3}{2}N_c ).
\label{coeffi3}
\end{eqnarray} 

From eqs.(\ref{coeffi2}) and (\ref{coeffi3}), we can extract
the relation such as
\begin{eqnarray}
{\beta}^{d}_0(N_f) &=& -2{\gamma}_0(N_f), 
\nonumber \\
{\beta}_0(N_f) &=& -2{\gamma}^{d}_0(N_f).
\label{nduality}
\end{eqnarray} 

This may be viewed as a new type of duality Oehme first
discovered in \cite{oeh}! Note here that the variable $N_f$
on both sides refers to matter fields with different quantum
numbers, i.e., one is electric, the other is magnetic.
Eq.(\ref{nduality}) might be also interpreted as follows:
originally in Seiberg's duality argument the interval
(\ref{conformal}) has been determined from the requirement
of asymptotic freedom in both electric and magnetic theories,
which means ${\beta}_0 < 0$ and ${\beta}^{d}_0 < 0$.
We believe here, however, that we could find a set of parameters
describing the interval (\ref{conformal}) 
{\it only} from those of the
original electric theory. Then anomalous dimension $\gamma_0$
might be a candidate. 
Below we shall give the same
considerations to various gauge theories with various matter
representations to check whether a new type of duality
(\ref{nduality}) proposed by Oehme is satisfied.

\vskip0.2truein  
{\bf (II) $G = SO(N_c)$ with fundamental matters of $N_f$ flavors}

Next we shall investigate the case of $SO(N_c)$ gauge theory
with massless superfields of $N_f$ flavors in the fundamental
representation. In this case ${\beta}_0$ and $\gamma_0$ are
given by \cite{pes}
\begin{eqnarray}
{\beta}_0 &=& -\frac{1}{16{\pi}^2}[3(N_c-2) - N_f], 
\nonumber \\
{\gamma}_0 &=& -\frac{1}{16{\pi}^2}[\frac{3}{2}(N_c-2) - N_f].
\label{coeffi4}
\end{eqnarray} 
Requiring ${\beta}_0 < 0$ and $\gamma_0 > 0$, we obtain
\begin{equation}
\frac{3}{2}(N_c - 2) < N_f < 3(N_c - 2).
\label{conformal2}
\end{equation}
This is nothing but the region corresponding to the conformal
window discussed in \cite{intri}. Then corresponding dual
magnetic theory is $G_{dual} = SO(N_f-N_c+4)$ gauge theory
with fundamental matter fields of $N_f$ flavors. Dual one-loop
coefficients become as
\begin{eqnarray}
{\beta}^{d}_0 &=& -\frac{1}{16{\pi}^2}[2N_f-3(N_c-2)], 
\nonumber \\
{\gamma}^{d}_0 &=&
             -\frac{1}{16{\pi}^2}[\frac{1}{2} N_f-\frac{3}{2}(N_c-2)].
\label{coeffi5}
\end{eqnarray} 
Thus we can conclude
\begin{eqnarray}
{\beta}^{d}_0(N_f) &=& -2{\gamma}_0(N_f), 
\nonumber \\
{\beta}_0(N_f) &=& -2{\gamma}^{d}_0(N_f).
\label{nduality2}
\end{eqnarray} 
In this case we find Oehme's duality also holds.

\vskip0.2truein
{\bf (III) $G=Sp(2N_c)$ with fundamental 
  matter fields of $N_f$ flavors}

This theory has also the magnetic description in a certain values of
$N_c$ and $N_f$. Dual gauge theory is $G_{dual}$ = $Sp(2N_f-2N_c-4)$
with fundamental matter superfields of 
$N_f$ flavors \cite{sp}. The one-loop
coefficients of both theories are given as follow:
\be G =
Sp(2N_c) ~~~\lq \lq electric"~~~ N = 1 ~~ SUSY \cr
{}~~~~~~~~~~~~~~~~~~~~~~~~~~~~~~~~~~~~~~~~~~~~\cr \beta_0
~~~=~~~-\frac{1}{16\pi^2}[3(2N_c+2) ~-~ 2N_f ] \cr \gamma_{0}
~~~=~~~-\frac{1}{16\pi^2}[\frac{3}{2}(2N_c+2) ~-~ 2N_f] ~~,
\label{sp1}
\en and \be G = Sp(2N_f-2N_c-4) 
~~~\lq \lq magnetic"~~~N = 1 ~~ SUSY \cr
{}~~~~~~~~~~~~~~~~~~~~~~~~~~~~~~~~~~~~~~~~~~~~~~~\cr \beta^d_{0}
~~~=~~~-\frac{1}{16\pi^2}[4N_f ~-~ 3(2N_c+2)] \cr \gamma^d_{0}
~~~=~~~-\frac{1}{16\pi^2}[N_f ~-~ \frac{3}{2}(2N_c+2)] ~.
\label{sp2}
\en
From (\ref{sp1}) and (\ref{sp2}), we get the results
${\beta}^{d}_0 = -2{\gamma}_0$
and ${\beta}_0 = -2{\gamma}^{d}_0$ in this case, too.

\vskip0.2truein
{\bf (IV) $G=G_2$ with fundamental 
matter fields of $N_f$ flavors}

This case may be a little non-trivial.
Original electric gauge theory has $G=G_2$ with
$N_f$ flavors of massless superfields in the fundamental
representation. In this case \cite{pesan},
\begin{eqnarray}
{\beta}_0 &=& -\frac{1}{16{\pi}^2}(12 - N_f), 
\nonumber \\
{\gamma}_0 &=& -\frac{1}{16{\pi}^2}(6 - N_f).
\label{coeffi6}
\end{eqnarray} 
Then the interval $6 < N_f < 12$ corresponds to interacting
non-Abelian Coulomb phase as discussed in \cite{pesan}.
There exists the magnetic description in this interval.
Dual magnetic gauge theory has $G_{dual}=SU(N_f-4)$ with
fundamental $N_f$ massless matter fields. The one-loop coefficients
are given by
\begin{eqnarray}
{\beta}^{d}_0 &=& -\frac{1}{16{\pi}^2}[3(N_f-4) - N_f]
               = -\frac{1}{16{\pi}^2}(2N_f-12), 
\nonumber \\
{\gamma}^{d}_0 &=& -\frac{1}{16{\pi}^2}[\frac{3}{2}(N_f-4)-N_f]
                = -\frac{1}{16{\pi}^2}(\frac{1}{2}N_f-6).
\label{coeffi7}
\end{eqnarray} 
Compared (\ref{coeffi6}) with (\ref{coeffi7}),
we can come to the conclusion that Oehme's duality is satisfied.
This seems to be rather non-trivial check for this duality.

\vskip0.2truein
{\bf (V) $G=SU(N_c), SO(N_c), Sp(2N_c)$ 
with massless adjoint matter}

After a short time of the discovery of original non-Abelian dualities
by Seiberg et al., Kutasov extended Seiberg's argument to 
$SU(N_c)$ gauge theory
including not only fundamental but also adjoint matter superfield
\cite{kut}.
In this case 
the dual magnetic theory becomes an $SU(2N_f-N_c)$ gauge
theory with $N_f$ massless matter fields in the fundamental
representation and an adjoint field and a certain number of singlet
massless matter fields. Then the one-loop coefficients in both
theories are of the form
\begin{eqnarray}
{\beta}_0 &=& -\frac{1}{16{\pi}^2}(2N_c - N_f), 
\nonumber \\
{\gamma}_0 &=& -\frac{1}{16{\pi}^2}(\frac{1}{2}N_c - N_f).
\label{coeffi8}
\end{eqnarray} 
and
\begin{eqnarray}
{\beta}^{d}_0 &=& -\frac{1}{16{\pi}^2}(2N_f-3N_c), 
\nonumber \\
{\gamma}^{d}_0 &=& -\frac{1}{16{\pi}^2}(-\frac{1}{2} N_c).
\label{coeffi9}
\end{eqnarray} 
We find in this case Oehme's duality checked above does not hold!

Similar analysis can be done for $SO(N_c)$ and $Sp(2N_c)$ gauge
groups with adjoint fields \cite{adj}
and leads us to the same results, i.e.,
Oehme's duality condition is not also satisfied in these cases.

\vskip0.2truein
{\bf (VI) $G=SU(N_c)$ with an antisymmetric tensor field}

As the final example, we investigate the theory of $G=SU(N_c)$
with an antisymmetric tensor field originally discussed in 
\cite{pou}. Remarkably the dual magnetic gauge group does not
become simple but product one in this case. And this theory
is also attractive as a model of the supersymmetry breaking 
\cite{pou}.

 The one-loop coefficients in the original electric theory are 
\begin{eqnarray}
{\beta}_0 &=& -\frac{1}{16{\pi}^2}
(2N_c -N_f + 3), 
\nonumber \\
{\gamma}_0 &=& -\frac{1}{16{\pi}^2}(\frac{1}{2}N_c - N_f + 3).
\label{coeffi10}
\end{eqnarray} 
For $N_f > 5$, dual magnetic gauge theory exists. It is
represented as the product gauge group $SU(N_f-3) \otimes Sp(N_f-4)$
with five species of dual quark superfields: a field transforming
as a fundamental under both groups, a conjugate antisymmetric 
tensor, a fundamental and $N_f$ antifundamentals of $SU(N_f-3)$
and also $N_c+N_f-4$ fundamentals of $Sp(N_f-4)$.

Then one-loop coefficients in each group are given as follows:
\begin{eqnarray}
{\beta}^{d}_0 &=& -\frac{1}{16{\pi}^2}(2N_f - \frac{9}{2}), 
\nonumber \\
{\gamma}^{d}_0 &=& -\frac{1}{16{\pi}^2}
(\frac{1}{2}N_f - 3)
\qquad \qquad in \quad SU(N_f-3).
\label{coeffi11}
\end{eqnarray} 
and
\begin{eqnarray}
{\beta}^{d}_0 &=& -\frac{1}{16{\pi}^2}
(\frac{5}{2}N_f - \frac{1}{2}N_c - \frac{9}{2}), 
\nonumber \\
{\gamma}^{d}_0 &=& -\frac{1}{16{\pi}^2}
(N_f - \frac{1}{2}N_c - \frac{3}{2})
\qquad \qquad in \quad Sp(N_f-4).
\label{coeffi12}
\end{eqnarray} 

Thus we conclude from these equations that Oehme's duality
relation also does not hold in this final example.

\section{Discussions}

In this paper, we discussed superconvergence relation 
following to the original work by Oehme and Zimmermann and
then investigated whether a new type of duality recently 
proposed by Oehme in $N$=1 supersymmetric gauge theory
also holds in other gauge theories which have different gauge
groups and different matter contents.

 As a result, we found that in the cases of gauge groups with
matter fields 
{\it only} in the fundamental representation, Oehme's
duality was satisfied while in those of the theories including
adjoint or antisymmetric tensor matter fields, it was not.
The reason may be as follows:  
when we add an adjoint field to a theory,
one can add a superpotential.
The model without a superpotential will
presumably flow in the infrared to a fixed point,
while adding the superpotential drives the system to
a new fixed point \cite{kut}. Kutasov's duality holds only
in the model with a superpotential.
On the other hand, in superconvergence argument
in which we evaluate $\beta_0$ and $\gamma_0$
in equation (\ref{perturb}),
we cannot distinguish a theory with
a superpotential from the one without it
because the contribution of the superpotential 
to $\beta$-function is higher
order one. Actually it appears not in 
$\beta_0$ but in $\beta_1$ in eq.(\ref{perturb}).
 

%
%
%
%
%
%
%
%

 Also we have to consider the problem of the gauge dependence
of our method \cite{on}. 
Originally gluon propagator is unphysical, 
because it depends on the gauge parameter $\alpha$. 
In this paper, we chose Landau gauge ($\alpha=0$) 
in convenience. Certainly the value of $\beta_0$ does not
depend on the specific gauge choice
while that of $\gamma_0$ does.
Therefore even if we obtain the result $\gamma_0 < 0$ 
in Landau gauge, $\gamma_0 > 0$ 
might be realized when we move to other gauges.
If so, our superconvergence argument based on the value
of $\gamma_0$  might not be believed.
We shall reconsider this point in another occasion.
\vskip0.5truein
\centerline{{\it ACKNOWLEDGMENTS}}
We would like to thank to our colleagues
in Kobe University for valuable comments and discussions. 
We are especially grateful to M. Sakamoto for his valuable
comments and various discussions about duality.
And we would like to
appreciate Prof. N. Nakanishi and 
Prof. K. Nishijima for their kindness.

\newpage

{\bf Appendices}
\vskip0.3truein
{\bf Appendix A: Proof of equation (\ref{sol})}

In Appendix A, we prove the equation (\ref{sol}).
Starting point is 
\begin{equation}
u\frac{\partial R(u,g)}{\partial u}
= \beta (g^2)\frac{\partial R(u,g)}{\partial g^2}
 +\gamma (g^2)R(u,g).
\label{cgeq2}
\end{equation}
Here we define $\tilde{R}$ as follows:
\begin{equation}
\tilde{R} \equiv \tilde{R}(Q^2(u,g),g)= R(u,g).
\label{tilder}
\end{equation}
Then we find (\ref{cgeq}) can be rewritten as
\begin{equation}
u\frac{\partial Q^2}{\partial u}|_g
\frac{\partial \tilde{R}}{\partial Q^2}|_g
= \beta \bigl( \frac{\partial \tilde{R}}{\partial g^2}|_u
+\frac{\partial Q^2}{\partial g^2}|_u
 \frac{\partial \tilde{R}}{\partial Q^2}|_u \bigr)
 +\gamma \tilde{R}.
\label{cgeq3}
\end{equation}
On the other hand, the effective coupling constant $Q$ is through
the equation (\ref{qeq}), i.e.,
\begin{equation}
u\frac{\partial Q^2}{\partial u}|_g
= \beta (g^2)\frac{\partial Q^2}{\partial g^2}|_u.
\label{qeq2}
\end{equation}
Substituting this relation (\ref{qeq2}) into (\ref{cgeq2}),
we will obtain
\begin{equation}
0 = \beta(g^2)\frac{\partial \tilde{R}(Q^2,g^2)}{\partial g^2}
   +\gamma(g^2) \tilde{R}(Q^2,g^2),
\label{cgeq4}
\end{equation}
and we can easily solve this equation as the following form
\begin{equation}
\tilde{R}(Q^2,g^2) = \tilde{R}(Q^2,Q^2)
exp \bigl[ \int^{Q^2}_{g^2}dx \frac{\gamma(x)}{\beta(x)} \bigr].
\label{sol2}
\end{equation}
Moreover, by definition,  
\begin{eqnarray}
R(1,g^2) &=& \tilde{R}\bigl( Q^2(1,g^2),g^2 \bigr), \\ \nonumber
Q^2(1,g^2) &=& g^2.
\label{bydef}
\end{eqnarray}
From these equations we obtain $\tilde{R}\bigl( Q^2(1,g^2),g^2 \bigr)
=R(1,Q^2)$ and finally we have the result as
\begin{equation}
R(\frac{k^2}{\mu^2},g^2) = R(1,Q)
exp \bigl[ \int^{Q^2}_{g^2}dx \frac{\gamma(x)}{\beta(x)} \bigr].
\qquad \qquad Q.E.D.
\label{sol3}
\end{equation}
\vskip0.5truein
{\bf Appendix B: Proof of (13) with (14)}

In Appendix B, we evaluate eq.(\ref{sol}) proved in the Appendix A
at large momentum and show that (13) is satisfied. 
For that purpose we first separate the integrand of (\ref{sol})
in the following form
\begin{equation}
\gamma(x)\beta^{-1}(x) = \frac{\gamma_0}{\beta_0 x}+\tau(x),
\label{tau}
\end{equation}
where $\tau(x)$ is corresponding to the regular part of
$\gamma(x)\beta^{-1}(x)$ at nearly $x=0$. Note here that
at large momentum, the functions $\beta(x)$ and $\gamma(x)$
are given by (\ref{perturb}) respectively 
in which $g^2$ is replaced by $x$.

Then it is not difficult to show (13) and (14). In fact
(here we put $R(1,Q)=1$),
\begin{eqnarray}
R(\frac{k^2}{\mu^2},g^2) &=&  
exp \bigl[ \int^{Q^2}_{g^2}dx \frac{\gamma_0}{\beta_0 x} \bigr]
exp \bigl[ \int^{Q^2}_{g^2}dx \tau(x) \bigr], \\ \nonumber
&=& \bigl( \frac{Q^2}{g^2} \bigr)^{\frac{\gamma_0}{\beta_0}}
   exp \bigl[ \int^{Q^2}_{g^2}dx \tau(x) \bigr], \\ \nonumber
&\approx& \bigl[g^2|\beta_0|
        \ln(\frac{k^2}{\mu^2}) \bigr]^{-\frac{\gamma_0}{\beta_0}}
         exp \bigl[ \int^{Q^2}_{g^2}dx \tau(x) \bigr], \\ \nonumber
&\equiv& C_V(\ln\frac{k^2}{\mu^2})^{-\frac{\gamma_0}{\beta_0}}, 
\label{perturb2}
\end{eqnarray}
where
\begin{equation}
C_V = (g^2|\beta_0|)^{-\frac{\gamma_0}{\beta_0}}
      exp \bigl[ \int^{Q^2}_{g^2}dx \tau(x) \bigr].
      \qquad \qquad Q.E.D.
\label{cv}
\end{equation}
\vskip0.5truein
{\bf Appendix C: Proof of 
Superconvergence Relation (\ref{screlation})}

In this Appendix, we prove the equation (\ref{screlation}), the 
superconvergence relation which is main subject of this paper.
To do so, let us use (\ref{asymp2}), (\ref{lehmann}) and
(\ref{spectral2}) and combine them. 

First multiplying (\ref{lehmann}) into $k^2$
\begin{equation}
k^2 D(k^2) = \int^{\infty}_{0}dm^2 \frac{k^2\rho(m^2)}{m^2 - k^2}.
\label{lehmann2}
\end{equation}
Then we can show the following results:
\begin{eqnarray}
& &\lim_{k^2 \to -\infty}(l.h.s. of eq.(\ref{lehmann2}))
 \\ \nonumber
&=& \lim_{k^2 \to -\infty}k^2 D(k^2)
= \lim_{k^2 \to -\infty}k^2 D_{asymp}(k^2) \\ \nonumber
&=&  \lim_{k^2 \to -\infty}
    C_V(\ln\frac{k^2}{\mu^2})^{-\frac{\gamma_0}{\beta_0}}=0
   \qquad \qquad for \quad \frac{\gamma_0}{\beta_0}>0, 
\label{lehmann3}
\end{eqnarray}
on the other hand,
\begin{eqnarray}
& & \lim_{k^2 \to -\infty}(r.h.s. of eq.(\ref{lehmann2}))
\\ \nonumber
&=& \lim_{k^2 \to -\infty}
\int^{\infty}_{0}dm^2 \frac{k^2\rho(m^2)}{m^2 - k^2} \\ \nonumber
&=& \lim_{k^2 \to -\infty}\int^{\infty}_{0}dm^2 
\frac{-(m^2-k^2)\rho(m^2)+m^2\rho(m^2)}{m^2 - k^2} \\ \nonumber
&=& -\int^{\infty}_{0}dm^2\rho(m^2)
    +\lim_{k^2 \to -\infty}
\int^{\infty}_{0}dm^2\frac{m^2\rho(m^2)}{m^2 - k^2}.
\label{lehmann4}
\end{eqnarray}
Here we compute more in detail the integration in the last column:
\begin{eqnarray}
& &\lim_{k^2 \to -\infty}
\int^{\infty}_{0}dm^2\frac{m^2\rho(m^2)}{m^2 - k^2} \\ \nonumber
&=&\lim_{k^2 \to -\infty}
\Bigl( \int^{\Lambda^2}_{0}dm^2 \frac{m^2\rho(m^2)}{m^2 - k^2}
 + \int^{\infty}_{\Lambda^2}dm^2 \frac{m^2\rho(m^2)}{m^2 - k^2}
\Bigr), \\ \nonumber
&=& 0 + \lim_{k^2 \to -\infty}
\int^{\infty}_{\Lambda^2}dm^2 \frac{m^2\rho_{asymp}(m^2)}{m^2 - k^2},
\\ \nonumber
&=& \lim_{k^2 \to -\infty}\int^{\infty}_{\Lambda^2}dm^2
\frac{-\frac{\gamma_0}{\beta_0}C_V
\bigl( \ln\frac{m^2}{\mu^2} \bigr) ^{-\frac{\gamma_0}{\beta_0}-1}}
{m^2 - k^2} = 0.
\label{lehmann5}
\end{eqnarray}
Compared with both sides of (\ref{lehmann2}), we have the result
that
\begin{equation}
\int^{\infty}_0 dm^2 \rho(m^2) = 0.  \qquad \qquad Q.E.D.
\label{sc'}
\end{equation}

\vskip0.5truein

\end{document}